\documentclass[10pt, onecolumn]{IEEEtran}

\pdfoutput=1

\usepackage{hyperref}
\usepackage{color}
\usepackage{amsmath}
\usepackage{algorithm}
\usepackage{algorithmic}
\usepackage{array}
\usepackage{mdwmath}
\usepackage{eqparbox}
\usepackage{float}
\usepackage[font=footnotesize]{subfig}
\usepackage{fixltx2e}
\usepackage{stfloats}
\usepackage{amsfonts}
\usepackage{cite}
\usepackage{amsmath}
\usepackage{bbm}
\usepackage{amssymb}
\usepackage{mathrsfs}
\usepackage{amsbsy}	
\usepackage{graphicx}

\floatstyle{ruled}
\newfloat{algorithm}{tbp}{loa}
\floatname{algorithm}{Algorithm}

\long\def\hide#1{}

\newcommand{\droped}[1]{{\color{blue} \sout{}}}

\def\ba{\begin{array}}
\def\ea{\end{array}}
\newcommand{\beq}{\begin{equation}}
\newcommand{\eeq}{\end{equation}}
\newcommand{\bq}{\begin{eqnarray}}
\newcommand{\eq}{\end{eqnarray}}
\newcommand{\bqn}{\begin{eqnarray*}}
\newcommand{\eqn}{\end{eqnarray*}}
\newcommand{\bee}{\begin{enumerate}}
\newcommand{\eee}{\end{enumerate}}
\newcommand{\bi}{\begin{itemize}}
\newcommand{\ei}{\end{itemize}}

\begin{document}

\title{Benefit of Multipath TCP on the Stability of Network}

\author{Xiuli Hu, Pangbei Hong and Bing Li \\
School of Computer Science, Hubei University of Technology}

\maketitle

\begin{abstract}
\bf{
Multipath-TCP receives a lot of attention recently and can potentially improve quality of service for both private and commercial users. It leverages the multiple available paths and send packets through all the available paths. The growing of Mutipath TCP has received a growing interest from both researchers who publish a growing number of articles on the topic and the vendors since Apple has decided to use Multipath TCP on its smartphones
and tablets to support the Siri voice recognition application. In this paper, we study the performance of Multipath TCP from its impact on the stability of the network. In particular, we study three scenarios, Internet, which is the largest networks and involves heterogeneous traffic, data center, which is smaller but has different traffic patterns compared with Internet scale network and wireless network, whose energy consumption also needs to be considered. Our study shows that stability is affected but not serisouly for Internet and wireless network, but datacenter network stability is seriously affected due to its bursty traffic pattern. 
}
\end{abstract}

\section{Introduction}

MultiPath TCP (MPTCP) is an effort towards enabling the simultaneous use of several IP-addresses/interfaces by a modification of TCP that presents a regular TCP interface to applications, while in fact spreading data across several available interfaces. It has various benefits including better resource utilization, better throughput and smoother reaction to failures. It's enabled by the fact that many devices today can connect to the network using different IP address, {e.g.} for mobile devices, they have 4G/LTE access while can also connect to WiFi hot spot, in datacenter, it can connect to the switch using various connections. 

Multipath TCP (MP-TCP) is a TCP extension that allows a TCP connection to stripe traffic over multiple interfaces. It is being standardized by the IETF \cite{IETF} and they are in active calling for proposal. Various aspects on MP-TCP have been investigated in the literature. In some of the early works \cite{FPKelly_MTCP,RSrikant}, they analyze the stability of using MPTCP in general network using fluid model and show that the aggregate throughput of a source converges to an equilibrium even when there is feedback delay. 
Later on, algorithms based on heuristic are developed that archives better stability performance than the early works \cite{Damon,ford2011architectural,raiciu2011improving}. However, there is no systematic understanding of the performance of those algorithms and it's generally unknown how the algorithm performs when the network becomes real and complex. Analysis based on some assumptions are derived in \cite{raiciu2012hard,Peng2014MultipathTCP}.

All the above works focus on general Internet network. Datacenter network, which is both smaller in scale and more bursty in traffic, receives much more attention recently due to the dramatic penetration of data center across the globe. In \cite{al2008scalable}, the authors propose a multipath routing algorithm for data center network. Multipath routing algorithm shares similarities with Multipath TCP algorithms but they provide multiple accesses from different network layer. In \cite{raiciu2011improving}, the first mutlipath TCP algorithm is proposed and extensively tested in the real data center network and is shown to achieve higher throughput and applying regular TCP algorithm directly into the data center. Thus, a lot of attentions are paid, including exepriment from different aspects, \cite{hesmans2015first,li2014tolerating}. Recently, a new Mutipath TCP algorithm based on different network architecture has been proposed that achieves both stability and throughput improvement \cite{greenberg2008towards}.

Wireless Multipath TCP, which is another big application of Multipath TCP, also receives a lot of attention recently. Different from existing 
work on wireline network, energy efficiency is a very important performance metric for mobile devices including cellphones and tablets. Thus, algorithms based on improving both energy efficiency and throughput performance for mobile devices have been studied extensively over the last decades, including \cite{wirelessall,wireless1,EWTCP,2011mobility,lim2014green,peng2014energy,chen2013measurement,zhou2014multipath,sridharan2014multi}. Various advantages of MP-TCP for mobile devices have been reported. In \cite{Damon}, the authors show that the throughput can be improved by using cellular and wirelessLAN network. Mobility is also shown to achieve higher througput for users with Multipath TCP \cite{2011mobility} because smooth switch between 4G and WiFi can be achieved.Thus, MP-TCP is promising in providing reliable and efficient connections for mobile devices by using connection diversity and resource pooling under dynamic environments. Please refer to more discussion on this pointer in \cite{perrucci2011survey,pqymptcp}.

Many energy saving mechanisms have been proposed to reduce energy consumption when MP-TCP is implemented in mobile devices. In \cite{pluntke2011saving}, a scheduler based on stochastic process is proposed. In \cite{2011mobility}, a mechanism that selects the most energy efficient path based on periodic probing of all paths are proposed. Moreover, since energy cost is the criterion for path selection, the performance would be affected if the energy efficient path turns out to be the most congested path. In \cite{peng2014energy}, an energy efficient Multipath TCP is proposed that intelligently identify good paths.

All the above work is based on loss based TCP algorithm, i.e. they reduce the window size when they observe packet loss on a particular paths. Delay based TCP algorithm, which adjusts the window size based on the delay in that path, is studied under Multipath TCP scenario, \cite{guo2014delay}.

\begin{figure}
\begin{center}
\includegraphics[width=0.4\textwidth]{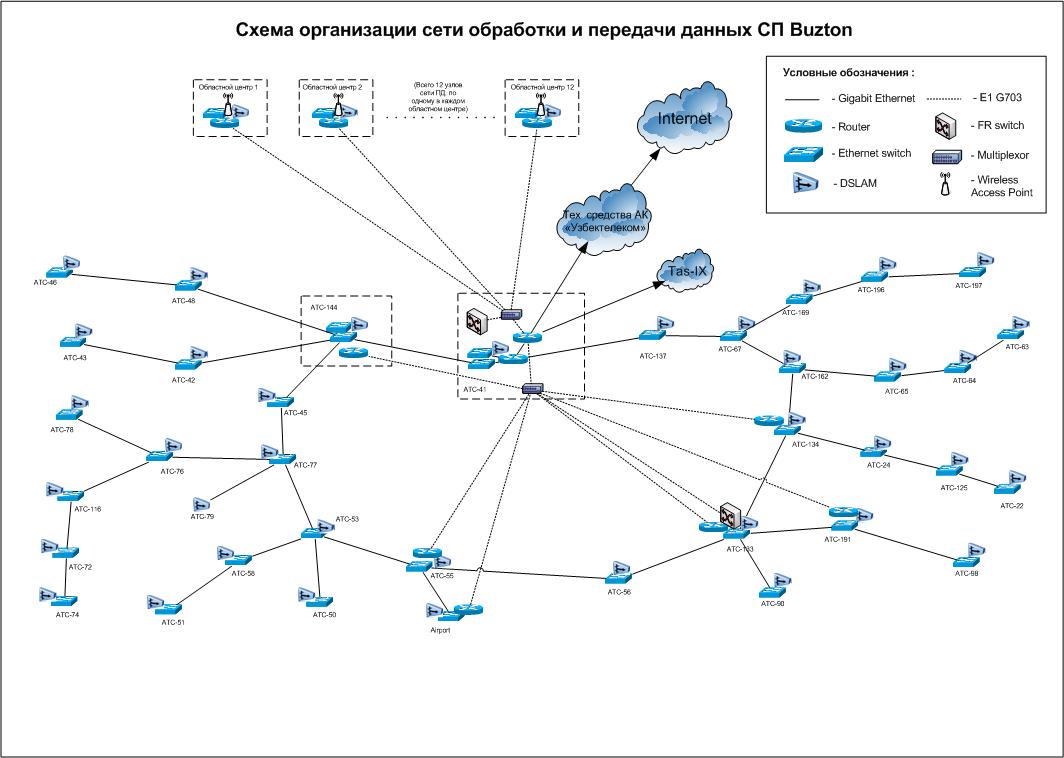}
\caption{Diagram of Internet Network. There are billions of nodes connecting to each other}
\label{default1}
\end{center}
\end{figure}

\begin{figure}
\begin{center}
\includegraphics[width=0.4\textwidth]{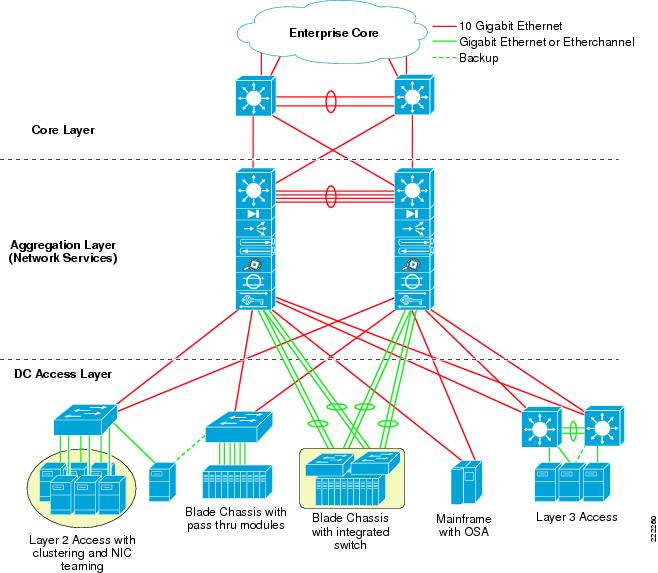}
\caption{Diagram of Data center network. There are usually three layers and most of the nodes stay in the bottom layer. They have multiple connections to the upper switch}
\label{default2}
\end{center}
\end{figure}

\begin{figure}
\begin{center}
\includegraphics[width=0.4\textwidth]{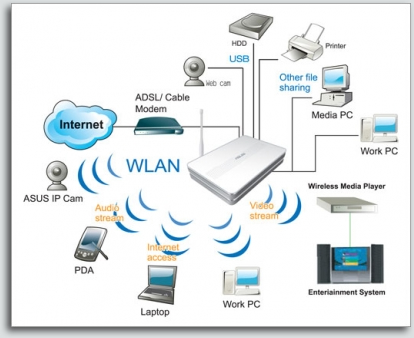}
\caption{Diagram of Wireless network.}
\label{default3}
\end{center}
\end{figure}

Our goal is to develop a holistic approach that can justify whether Multipath TCP is useful, i.e. whether it affects network stability under various scenairos, including
\begin{itemize}
\item Internet: Internet is a highly dynamic environment, where there are billions of users any time and is very complex and heterogeneous. (Fig. \ref{default1})
\item Data center: data center is smaller in scale compared with Internet. However, the environment is totally different from Internet due to mice traffic. (Fig. \ref{default2})
\item Wireless network, wireless network is further smaller in scale. However, they typically are higher in delay (round trip time) and incur high energy consumption, which is crucial to wireless mobile devices. Thus, it's crucial to achieve lower energy consumption, which may put the wireless network is an unstable state. (Fig. \ref{default3})
\end{itemize}

The paper is structured as below. In section II, we provide the model and notatios. In section III, we summarize the theory used extensively in Multipath TCP. In section IV, we analyze the stability using the model for various scenarios. The paper is concluded in section V.

\section{Problem formulation}

In this section, we introduce our notations and model and formulate the problem used for stability assessment. 

\subsection{Notations}
In this paper, we use the following mathematical notations throughout the paper:
\begin{itemize}
\item$r_p$: Rate of path $p$.
\item$r_x$: Total rate of a source $x$, which is assumed to have at least one paths.
\item $U_x(r_x)$: Utility function associated with a source $x$ with total rate $r_x$.
\item $\mathbb{R}^n$:           n dimension real number.
\item $\mathbb{C}^n$:           n dimension complex number.
\item $\|x\|$:          Euclidean distance of $x$, defined as
\begin{equation*}
\|x\|=\sqrt{\sum_{i=1}^{n}x_i^2}
\end{equation*}
represents the Euclidean length of a vector $x\in \mathbb{R}^n$, then $\|x+y\|\leq\|x\|+\|y\|$.
\end{itemize}

Let $\mathbb{R}^n$ denote the set of $n$ dimensional real number and $\mathbb{C}^n$ denote the set of $n$ dimensional complex number. $x\in\mathbb{R}^n$ means $x$ is a $n$ dimensional vector. Let

A function 
\begin{equation*}
U(x) \in \mathbb{R}\rightarrow \mathbb{R}
\end{equation*}
is defined as a {\bf Utility function} if $U(x)$ is a concave function of $x$.

A utility function that incorporates energy consumption is defined as 
\begin{equation}
U(x,e) \in \mathbb{R}^2\rightarrow \mathbb{R}
\end{equation}
defines users satisfaction of throughput at $x$ and energy consumption at $e$. 

A function 
\begin{equation*}
U(x) \in \mathbb{R}\rightarrow \mathbb{R}
\end{equation*}
is defined as a {\bf Utility function} if $U(x)$ is a concave function of $x$.

\subsection{Problem formulation}
In this section, we define how stability is assessed using a stability assessment model.

Let $x^*$ be the old equilibrium, and let $x^n$ be the new equilibrium when multipath TCP algorithm is used. Then stability is defined as the difference between these two points, i.e.
\bqn
\|x^n-x^*\|.
\eqn
And our objective is to minimize this distance, i.e.
\bqn
\min \quad \|x^n-x^*\|.
\eqn
There are various constraints we need to satisfy:
\begin{itemize}
\item Number of paths $N_x$ associated with source $x$: $N_x\geq 2$, i.e. each source has more than one path. This assumption is due to the fact that we assume each source has more than one path.
\item Throughput on paths $r_p$: $r_p\geq \epsilon>0$ for a given $\epsilon$. This make sure that each path is void to waste network resoruces.
\item The total traffic through a link should be smaller than its capacity: 
\begin{equation*}
\sum_{i\in y\cap R_p} r_i\leq Capacity_i
\end{equation*}
\item Traffic burdens control $b_p$ on each path $p$.
\bqn
\|b_p^*-b_p^n\|< \infty
\eqn
where $b_p^*$ is the old equilibrium and $b_p^n$ is the new equilibrium. 
\end{itemize}
\section{Analysis}

In this section, we will show assess the stability of different network scenarios based on the above formulation for all the three types of network. We show that accurately assess the performance us NP hard due to the exponential combinations of scenarios. 

\subsection{Interent}
Internet is as stable as the original cases. 
\subsection{Data center}
Data center is not stable due to the bustiness and it violates the bustiness constraint.
\subsection{Wireless Network}
Wireless network is similar to Internet and stable as it is. 


\section{Conclusion}
In this paper, we study the stability issue associated with a Multipath TCP algorithm. Multipath-TCP receives a lot of attention recently and can potentially improve quality of service for both private and commercial users. It leverages the multiple available paths and send packets through all the available paths. The growing of Mutipath TCP has received a growing interest from both researchers who publish a growing number of articles on the topic and the vendors since Apple has decided to use Multipath TCP on its smartphones
and tablets to support the Siri voice recognition application. In this paper, we study the performance of Multipath TCP from its impact on the stability of the network. In particular, we study three scenarios, Internet, which is the largest networks and involves heterogeneous traffic, data center, which is smaller but has different traffic patterns compared with Internet scale network and wireless network, whose energy consumption also needs to be considered. Our study shows that stability is affected but not serisouly for Internet and wireless network, but datacenter network stability is seriously affected due to its bursty traffic pattern. 

\bibliographystyle{ieeetr}
\bibliography{reference}

\begin{thebibliography}{10}

\bibitem{IETF}
A.~Ford, C.~Raiciu, M.~Handley, S.~Barre, and J.~Iyengar, ``Architectural
  guidelines for multipath tcp development,'' {\em RFC6182 (March 2011), www.
  ietf. ort/rfc/6182}, 2011.

\bibitem{FPKelly_MTCP}
F.~Kelly and T.~Voice, ``Stability of end-to-end algorithms for joint routing
  and rate control,'' {\em ACM SIGCOMM Computer Communication Review}, vol.~35,
  no.~2, pp.~5--12, 2005.

\bibitem{RSrikant}
H.~Han, S.~Shakkottai, C.~Hollot, R.~Srikant, and D.~Towsley, ``Overlay tcp for
  multi-path routing and congestion control,'' in {\em IMA Workshop on
  Measurements and Modeling of the Internet}, 2004.

\bibitem{Damon}
D.~Wischik, C.~Raiciu, A.~Greenhalgh, and M.~Handley, ``Design, implementation
  and evaluation of congestion control for multipath tcp,'' in {\em Proceedings
  of the 8th USENIX conference on Networked systems design and implementation},
  pp.~8--8, USENIX Association, 2011.

\bibitem{ford2011architectural}
A.~Ford, C.~Raiciu, M.~Handley, S.~Barre, J.~Iyengar, {\em et~al.},
  ``Architectural guidelines for multipath tcp development,'' {\em IETF,
  Informational RFC}, vol.~6182, pp.~2070--1721, 2011.

\bibitem{raiciu2011improving}
C.~Raiciu, S.~Barre, C.~Pluntke, A.~Greenhalgh, D.~Wischik, and M.~Handley,
  ``Improving datacenter performance and robustness with multipath tcp,'' {\em
  ACM SIGCOMM Computer Communication Review}, vol.~41, no.~4, pp.~266--277,
  2011.

\bibitem{raiciu2012hard}
C.~Raiciu, C.~Paasch, S.~Barre, A.~Ford, M.~Honda, F.~Duchene, O.~Bonaventure,
  M.~Handley, {\em et~al.}, ``How hard can it be? designing and implementing a
  deployable multipath tcp.,'' in {\em NSDI}, vol.~12, pp.~29--29, 2012.

\bibitem{Peng2014MultipathTCP}
Q.~Peng, A.~Walid, J.~Hwang, and S.~Low, ``Multipath tcp: Analysis, design and
  implementation,'' {\em Networking, IEEE/ACM Transactions on}, 2014.

\bibitem{al2008scalable}
M.~Al-Fares, A.~Loukissas, and A.~Vahdat, ``A scalable, commodity data center
  network architecture,'' {\em ACM SIGCOMM Computer Communication Review},
  vol.~38, no.~4, pp.~63--74, 2008.

\bibitem{hesmans2015first}
B.~Hesmans, H.~Tran-Viet, R.~Sadre, and O.~Bonaventure, ``A first look at real
  multipath tcp traffic,'' in {\em Traffic Monitoring and Analysis},
  pp.~233--246, Springer, 2015.

\bibitem{li2014tolerating}
M.~Li, A.~Lukyanenko, S.~Tarkoma, Y.~Cui, and A.~Yl{\"a}-J{\"a}{\"a}ski,
  ``Tolerating path heterogeneity in multipath tcp with bounded receive
  buffers,'' {\em Computer Networks}, vol.~64, pp.~1--14, 2014.

\bibitem{greenberg2008towards}
A.~Greenberg, P.~Lahiri, D.~A. Maltz, P.~Patel, and S.~Sengupta, ``Towards a
  next generation data center architecture: scalability and commoditization,''
  in {\em Proceedings of the ACM workshop on Programmable routers for
  extensible services of tomorrow}, pp.~57--62, ACM, 2008.

\bibitem{wirelessall}
K.-K. Yap, T.-Y. Huang, M.~Kobayashi, Y.~Yiakoumis, N.~McKeown, S.~Katti, and
  G.~Parulkar, ``Making use of all the networks around us: a case study in
  android,'' in {\em Proceedings of the 2012 ACM SIGCOMM workshop on Cellular
  networks: operations, challenges, and future design}, pp.~19--24, ACM, 2012.

\bibitem{wireless1}
Y.-C. Chen, E.~M. Nahum, R.~J. Gibbens, and D.~Towsley, ``Measuring cellular
  networks: Characterizing 3g, 4g, and path diversity,'' tech. rep., UMass
  Amherst Technical Report: UM-CS-2012-022.

\bibitem{EWTCP}
M.~Honda, Y.~Nishida, L.~Eggert, P.~Sarolahti, and H.~Tokuda, ``Multipath
  congestion control for shared bottleneck,'' in {\em Proc. PFLDNeT workshop},
  2009.

\bibitem{2011mobility}
C.~Raiciu, D.~Niculescu, M.~Bagnulo, and M.~J. Handley, ``Opportunistic
  mobility with multipath tcp,'' in {\em Proceedings of the sixth international
  workshop on MobiArch}, pp.~7--12, ACM, 2011.

\bibitem{lim2014green}
Y.-s. Lim, Y.-C. Chen, E.~M. Nahum, D.~Towsley, and R.~J. Gibbens, ``How green
  is multipath tcp for mobile devices,'' in {\em Proceedings of the 4th
  workshop on All things cellular: operations, applications, \& challenges},
  pp.~3--8, ACM, 2014.

\bibitem{peng2014energy}
Q.~Peng, M.~Chen, A.~Walid, and S.~Low, ``Energy efficient multipath tcp for
  mobile devices,'' in {\em Proceedings of the 15th ACM international symposium
  on Mobile ad hoc networking and computing}, pp.~257--266, ACM, 2014.

\bibitem{chen2013measurement}
Y.-C. Chen, Y.-s. Lim, R.~J. Gibbens, E.~M. Nahum, R.~Khalili, and D.~Towsley,
  ``A measurement-based study of multipath tcp performance over wireless
  networks,'' in {\em Proceedings of the 2013 conference on Internet
  measurement conference}, pp.~455--468, ACM, 2013.

\bibitem{zhou2014multipath}
D.~Zhou, W.~Song, P.~Wang, and W.~Zhuang, ``Multipath tcp for user cooperation
  in lte networks,'' {\em IEEE Network. http://cs. unb. ca/\~{}
  wsong/publications/journals/NET\_MPTCP\_Dizhi. pdf}, 2014.

\bibitem{sridharan2014multi}
A.~Sridharan, R.~K. Sinha, R.~Jana, B.~Han, K.~Ramakrishnan,
  N.~Shankaranarayanan, and I.~Broustis, ``Multi-path tcp: Boosting fairness in
  cellular networks,'' in {\em Network Protocols (ICNP), 2014 IEEE 22nd
  International Conference on}, pp.~275--280, IEEE, 2014.

\bibitem{perrucci2011survey}
G.~P. Perrucci, F.~H. Fitzek, and J.~Widmer, ``Survey on energy consumption
  entities on the smartphone platform,'' in {\em Vehicular Technology
  Conference (VTC Spring), 2011 IEEE 73rd}, pp.~1--6, IEEE, 2011.

\bibitem{pqymptcp}
Q.~Peng, A.~Walid, and S.~H. Low, ``Multipath tcp algorithms: theory and
  design,'' in {\em Proceedings of the ACM SIGMETRICS/international conference
  on Measurement and modeling of computer systems}, pp.~305--316, ACM, 2013.

\bibitem{pluntke2011saving}
C.~Pluntke, L.~Eggert, and N.~Kiukkonen, ``Saving mobile device energy with
  multipath tcp,'' in {\em Proceedings of the sixth international workshop on
  MobiArch}, pp.~1--6, ACM, 2011.

\bibitem{guo2014delay}
W.~Guo, J.~Huang, and Y.~Zhang, ``Delay-based congestion control for multipath
  tcp,'' {\em traffic}, vol.~7, no.~1, 2014.

\bibitem{FPKelly_TCP}
F.~P. Kelly, A.~K. Maulloo, and D.~K. Tan, ``Rate control for communication
  networks: shadow prices, proportional fairness and stability,'' {\em Journal
  of the Operational Research society}, vol.~49, no.~3, pp.~237--252, 1998.

\end{thebibliography}


\end{document}